# A Hybrid Simulation-based Duopoly Game Framework for Analysis of Supply Chain and Marketing Activities

D. Xu[1], C. Meng[1], Q. Zhang[1], P. Bhardwaj[2], and Y.-J. Son[1,*]

**Abstract** A hybrid simulation-based framework involving system dynamics and agent-based simulation is proposed to address duopoly game considering multiple strategic decision variables and rich payoff, which cannot be addressed by traditional approaches involving closed-form equations. While system dynamics models are used to represent integrated production, logistics, and pricing determination activities of duopoly companies, agent-based simulation is used to mimic enhanced consumer purchasing behavior considering advertisement, promotion effect, and acquaintance recommendation in the consumer social network. The payoff function of the duopoly companies is assumed to be the net profit based on the total revenue and various cost items such as raw material, production, transportation, inventory and backorder. A unique procedure is proposed to solve and analyze the proposed simulation-based game, where the procedural components include strategy refinement, data sampling, gaming solving, and performance evaluation. First, design of experiment and estimated conformational value of information techniques are employed for strategy refinement and data sampling, respectively. Game solving then focuses on pure strategy equilibriums, and performance evaluation addresses game stability, equilibrium strictness, and robustness. A hypothetical case scenario involving soft-drink duopoly on Coke and Pepsi is considered to illustrate and demonstrate the proposed approach. Final results include *P*-values of statistical tests, confidence intervals, and simulation steady state analysis for different pure equilibriums.

[1] D. Xu, C. Meng, Q. Zhang, Y.-J. Son (son@sie.arizona.edu)
Systems and Industrial Engineering, the University of Arizona, Tucson, AZ, USA

[2] P. Bhardwaj
Operational Decision Support Technology, Intel Corporation, Tempe, AZ, USA

# Contents

# 1. Introduction

*Duopoly games* have been extensively studied in the modern history of economics, where the market is primarily dominated by two major companies and they make fully rational decisions to reach the goals (e.g. maximize payoff). While the most widely used approaches to solve the duopoly game are based on Cournot model (Cournot, 1838) and Bertrand model (Bertrand, 1883), several major limitations of those models are that:

- the *payoff function* of each company is highly aggregated by closed-form mathematical equations;
- only single or limited decision variables (e.g. production quantity, product price) are considered for mathematical tractability;
- no randomness involved in the payoff formulation.

In real practice, however, competing companies have to make and update decisions periodically on various areas such as production, logistics and price across the entire supply chain based on dynamically changing market conditions, and these decisions interact with one another to achieve a high profit. Hence, a comprehensive modeling technique is desired to mimic the realistic processes in multiple areas mentioned above, so as to provide a highly accurate payoff as well as to enable analysis of the trade-offs among different strategies.

In this chapter, a hybrid simulation-based framework is proposed to address duopoly game under the scenario of product adoption process considering multiple decision variables and detailed payoffs. In the proposed hybrid simulation framework,

- *system dynamics* (SD) models are used for simulating the activities of duopoly companies on production, logistics, and price determination;
- *agent-based simulation* (ABS) is used for modeling consumer purchasing behaviors at the market side.

Figure 1 outlines the major components in an exemplary supply chain and consumer market. In the SD model, an integrated production-logistics model considering the material transformations and flows from suppliers to final customers is constructed for each duopoly company. The price determination process, which is also modeled in the SD simulation, represents how each company determines the product price and adjusts it over time due to the impacts of production and logistics. To this end, an enhanced consumer motivation function is developed based on various factors such as the effect of advertisement, the effect of promotion, the influences of customer acquaintance recommendation, and the price sensitivity in the consumer social network. The consumer motivation function is then incorporated into the ABS for mimicking the consumer purchasing behaviors, which is tightly coupled with the SD model for the duopoly companies.

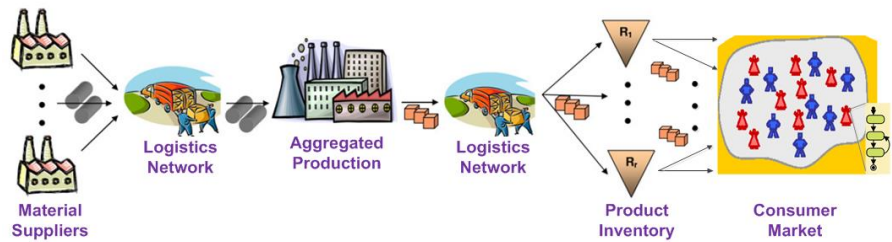

**Fig. 1** Exemplary supply chain and consumer market

Considering the *game strategy* for duopoly games, emphasis has been put in the following strategic areas (Min and Zhou, 2002; Hong *et al.*, 2008; Song and Jing, 2010) including production strategy (e.g. labor, raw material availability), logistics strategy (e.g. lead time, inventory coverage control), and marketing strategy (e.g. price determination, advertising, promotion). The strategic areas in the literature are coupled with the simulation model, so that any strategy changes can be reflected in simulation variables/parameters. In the game theory literature, each of the above strategic areas involves different decisions that are referred to as strategies. The payoff function of each dominated company is defined in terms of net profit, which is the difference between the revenue and various cost items such as production, logistics, transportation, and backorder. In the proposed work of this chapter, the objective for each duopoly company is assumed to maximize the net profit via the coordination of all the considered strategies.

In games involving a large number of strategies and data samples, conducting experiments including all the strategic decisions is computationally costly. In order to solve and analyze the *simulation-based game* in this work under limited computational resources, a novel procedure is proposed, where the procedural components include strategy refinement, data sampling, game solving, and performance evaluation. First, design of experiments technique used for strategy refinement and estimated conformational value of information (ECVI) technique used for data sampling are integrated for exploring the strategy space in the empirical game setting. Then, game solving for pure strategy equilibrium is applied to generate game equilibrium results, and performance evaluation approach is employed to assess various output criteria (e.g. equilibrium quality, stability, strictness and robustness). In the experiment section, a case with soft-drink duopoly game is considered to illustrate and demonstrate the framework.

Figure 2 depicts major components of the framework in this chapter:

- a hybrid simulation testbed of duopoly game with its profile set as inputs and payoff matrix as outputs (the upper part of Figure 2);
- a GSA procedure including strategy refinement, data sampling, game solving and performance evaluation (the lower part of Figure 2).

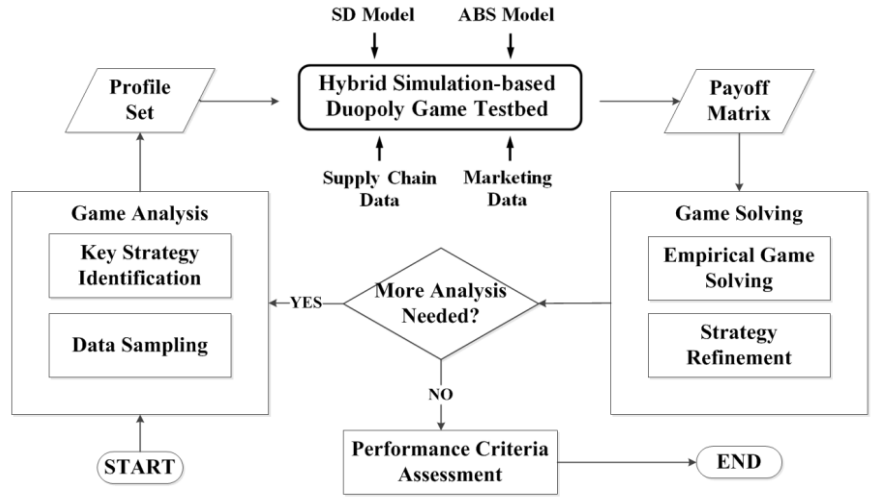

**Fig. 2** Proposed hybrid simulation framework with the GSA procedure

The major contributions of this chapter are summarized as follows:

1. A novel simulation-based *empirical game* platform is proposed, which overcomes the major drawbacks of closed-form mathematical equations in terms of modeling comprehensiveness;
2. A novel simulation-based *game solving and analysis* (GSA) procedure is proposed, which covers major topics in the field of game theory such as strategy refinement, data sampling, game solving, and performance evaluation.

In fact, the proposed simulation platform allows for accurate representation of the real world scenario, and it targets to address such game problem involving large strategy space and detailed/rich payoff function. Besides, the proposed platform is generic so that it can be re-used and further enhanced based on user requirements. The proposed GSA procedure is platform independent so that it can also be applied to resolve other similar simulation-based games.

The rest of this chapter is organized as follows. In Sect. 2, the literature works related to the proposed simulation-based game platform and the GSA procedure are summarized. In Sect. 3, the details of different modeling aspects (e.g. production, logistics, and marketing) that constitute the simulation-based game platform are provided, followed by the discussions of game strategies and payoff function. Sect. 4 discusses the motivation, objective of GSA, as well as its detailed procedure including strategy refinement, data sampling, game solving, and performance assessment. In Sect. 5, experiments are conducted and corresponding results are presented. Finally, conclusions and future directions are discussed in Sect. 6.

## 2. Background and Literature Survey

The game theoretic approach has been applied in the literature to address strategic decision problems in supply chain and marketing activities, where the studies mainly focused on the relationships between stakeholders within the supply chain system. For manufacturing strategy, Zhang and Huang (2010) investigated platform commonality and modularity strategies in a supply chain consisting of a single manufacturer and multiple cooperative suppliers. They derived the optimal ordering and pricing decisions for the two-moves dynamic game according to Nash's bargaining model, and developed an iterative algorithm to find the sub-game perfect equilibrium. They found that a supply chain with cooperative suppliers is more effective by using the lot-for-lot policy and more competitive by accommodating higher product variety. For logistics/inventory control strategy, Yu *et al.* (2006) studied Stackelberg game in a Vendor Managed Inventory (VMI) supply chain consisting one manufacturer as the leader and heterogonous retailers as followers. The research proposed a 5-step algorithm to reach the Stackelberg equilibrium and demonstrated 1) the significant influence of market-related parameters on manufacturer's and retailers' profit, 2) higher inventory cost does not necessarily lead to lowing retailers' profit and 3) game equilibrium benefits the manufacturer. The pricing and marketing strategies have been studied in an integrated manner in some literature works. Parlar and Wang (1994) studied the discounting strategy in a game involving one supplier with multiple homogeneous customers. They demonstrated that both seller and buyers can improve their own profit by using a proper discounting strategy. A similar game was also studied by Wang and Wu (2000). The difference was that the customers in this study were heterogeneous, and a price policy was proposed, where seller offers price discount based on the percentage increase from a buyer's quantity before discount. The proposed policy was demonstrated to provide benefits for venders compared with the one based on buyer's unit increase in order quantity. Esmaeili *et al.* (2009) proposed seller-buyer supply chain models considering pricing and marketing strategic decision variables such as price charged by seller to buyer, lot size, buyer's selling price, and marketing expenditure. Both cooperative and non-cooperative relationships between the seller and buyer were modeled assuming Seller-Stackelberg and Buyer-Stackelberg, respectively. The experiment results showed both optimal selling price and marketing expenditure were smaller in the cooperative game. While these works have provided guidance for addressing strategic decision making problems via a game theoretic approach, they faced limitations in efficiently obtaining accurate payoffs for a large strategy space under realistic case scenarios (e.g. duopoly company competition).

Most recently, simulation-based games have been employed to analyze complex interactions of players in the areas of supply chain (Collins *et al.*, 2004), combat (Poropudas and Virtanen, 2010), financial market (Mockus, 2010), subcontractor selection (Unsal and Taylor, 2011) and pedestrian behaviors (Asano *et*

*al.*, 2010). An advantage of this approach is that simulation is capable of modeling the detailed players' behaviors, their interactions as well as the external environment impacts. Hence, results from simulation are comprehensive and can be used for detailed analysis. To the best of our knowledge, although simulation-based game has been used for solving coordination problem within specific supply chain, a formal framework for solving integrated supply chain and its market competition game is not available in the literature. Next several paragraphs mainly survey the past research works that have formed a basis in this chapter in two aspects:

- SD and ABS modeling on supply chain and marketing activities, respectively;
- Approaches for empirical game analysis.

Concerning the simulation model for integrating the supply chain operations and marketing activities, different researchers have developed scenarios with distinct settings according to their own conveniences. To unify them under a coherent framework, the SD model in our work employs typical scenarios available in Sterman (2000) that involve labor utilization, raw material logistics, production process, and final production inventory control. However, necessary modifications have been made due to the duopoly game setting, and ABS integration for consumer purchasing behavior (see Sect. 3 for details). The consumer purchasing motivation and decision can be influenced by three factors (Kotler and Keller, 2007):

- personal (e.g. price sensitivity and quality sensitivity);
- social (e.g. adoption from word of mouth, follower tendency);
- psychological (e.g. perception and susceptibility to advertisement).

ABS can not only explore how and why consumers made the decision of purchasing certain products (North *et al.*, 2010), but also evaluate the overall system performance without sacrificing enough details on interdependency among company marketing behaviors. Previous researchers (Jager *et al.*, 1999; Adjali *et al.*, 2005; Yoshida *et al.*, 2007) have dealt with personal, social and psychological factors involving ABS technique. In this chapter, based on Zhang and Zhang (2007), an enhanced motivation function is proposed to incorporate the effects of advertisement, promotion from company, the influences of customer acquaintance recommendation, and price sensitivity in the consumer market. The consumer behavior modeled in ABS is coupled with the supply chain model to generate the market share and actual demand over time.

Previous literature works related to the simulation-based game analysis of this chapter are summarized in the following two paragraphs. A seminal research work in empirical game analysis is Wellman (2006), who decomposed the empirical game-theoretic analysis into three basic steps:

1. parameterize strategy space, which means to generate a set of candidate strategies from all available ones that are computationally intensive and costly ineffective to evaluate;
2. estimate the empirical game, which is aimed to construct empirical payoff matrix via simulation for the simplified game with the attention on the candidate strategies;
3. analyze (solve) the empirical game, and assess the solution quality with respect to the original game with full strategy sets.

For parameterizing strategy space, several baseline approaches are available in Wellman (2006) such as truthful revelation, myopic best response and game tree search. These methods have been applied in auction game (Reeves, 2005) and multi-player chess game (Kiekintveld *et al.*, 2006). For estimating the empirical game, two approaches exist in the literature, including direct estimation and regression. The first approach treats the observations as direct evidence for the payoffs of each player's strategy profile, while the idea of second method is to apply regression to fit an estimated payoff function over the entire profile space given the available data (Vorobeychik *et al.*, 2007). The goal of analyzing the empirical game is to find the pure and mixed strategy equilibrium firstly, and then to apply appropriate methods (e.g. statistical bounds) to gain insights into the original full game. Degree of game-theoretic stability is usually used to provide an ε-Nash equilibrium under this case.

Similar to our strategy refinement problem addressed in this chapter, Jordan *et al.* (2008) studied the profile selection problem with the objective of saving the computational costs for the promising equilibrium candidates. The authors studied different algorithms applicable to two different models: TABU best-response search (Sureka and Wurman, 2005) and minimum regret-first search (MRFS) for revealed-payoff; expected value of information (EVI) (Walsh *et al.*, 2003) and proposed information gain (IG) approach for noisy-payoff models. Later on, Jordan *et al.* (2010) solved a special case of the profile selection problem to determine an optimal simulation sequence of strategy sets. The paper also clarified the differences between the profile selection problem (Jordan *et al.*, 2008) and strategy exploration problem. Then, different exploration policies including random policy (RND), improving deviation only policy (DEV), best response policy (BR), softmax policy (ST) were discussed, followed by the experiments to compare their performances under different scenarios. For the sampling approach, Walsh *et al.* (2003) referred to the large/infinite number of strategies in the populated strategy space as heuristic strategies, and proposed two information theoretic approaches (i.e. EVI and ECVI approaches) to compute the additional sampling number for each experimental step. The paper demonstrated that ECVI approach converged faster than EVI given the same number of samples, and they both outperformed the uniform sampling approach. As pointed out in these literature works, when dealing with a large game strategy space, strategy exploration/refinement and data sampling are always the dominant costs for solving and analyzing the game, which constitute the major motivation for the development of the proposed GSA procedure in Sect. 4.

## 3. Hybrid Simulation-based Testbed for Duopoly Game Modeling

In this section, two major functional components constitute the simulation testbed: supply chain and marketing. The supply chain operations are modeled in SD, and marketing activities with its impact to the consumer behavior are modeled in ABS. The supplying process at the upstream is responsible for providing raw materials to the manufacturer. Production at the manufacturer begins when both raw material and labor are available. Inventories are kept along the supply chain to satisfy the customer orders at the downstream, and backorder is considered when the demand can't be fulfilled. The product price is also determined in the SD model, and it is impacted by the competitive product in the market, and production-logistics activities of its own company. Consumer purchasing behaviors are represented in the ABS model, which are highly related to the companies' market share and profit. All these modeling details are presented in the rest of this section.

### 3.1 System Dynamics for Modeling Production-Logistics Activities

Figures 3 and 4 are the snapshots for the production and logistics modules in the SD model, respectively, where equations from (1)-(21) represent underlying mathematical models and Table 1 provides nomenclatures for variables and parameters used in those equations. The concepts behind the SD model developed in this chapter are based on Sterman (2000) and Venkateswaran and Son (2007), with the enhancements and customizations made for our study. The major customizations/enhancements include:

- Duopoly game setting for our scenario;
- Interaction with the marketing module in ABS model;
- Incorporation of historical values via exponential moving average for adjusted production, inventory, labor, and vacancy;
- Incorporation of variations in demand, production, inventory, labor availability.

The entire production process has been aggregated into one stock in the SD model (see Eq. (1)). One assumption made when constructing these equations is that we treat the time as discrete variable, while in the SD model the corresponding variables change continuously. The adjusted WIP and production amounts are calculated via *exponential moving average* (smoothing) as shown in Eq. (2) and Eq. (3). As it is an order-driven inventory control and production system, the desired amount of WIP is calculated by multiplying the total of adjusted production amount and customer order rate with the manufacturing cycle time plus the variations (see Eq. (4)); the desired amount of production begin rate is calculated by summing up the adjusted WIP amount, adjusted production amount, customer order rate and the variations (see Eq. (5)).

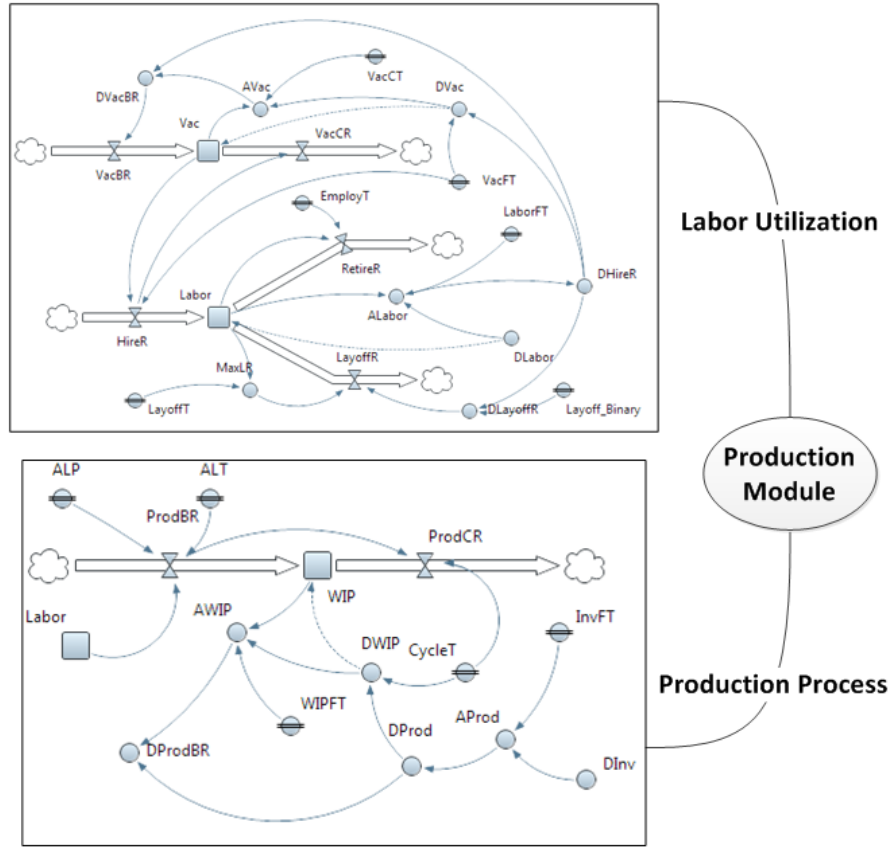

**Fig. 3** Production module in the simulation-based game testbed (customized from Sterman (2000))

In the ideal case, the actual production begin rate is equal to the desired production begin rate; however, it is always constrained by two other factors: *workforce availability* and *raw material availability* (see Eq. (6)). The availability of raw material is determined by the upstream supplier, of which the modeling is analogous to the logistics module of the finished goods (discussed later in this section). The labor changing process (e.g. vacancy creation and fulfillment) will be discussed in the next paragraph. The actual production begin rate equals to the minimal one (bottleneck) of the workforce, raw material amount, and desired production begin rate. The production cost is tightly related to the product price, which will be discussed later in this section.

One factor that influences the production plan is the labor availability. The labor is represented in one stock, and the labor vacancy rate is captured in another stock. The equations for calculating these two stock values are shown in Eq. (7) and Eq. (8). Hiring rate, retiring rate, and layoff rate are explicitly modeled in the SD model via Eq. (9), Eq. (10), and Eq. (11), respectively. These three rates are the major variables for deciding the labor availability, and a variable called vacancy begin rate will be increased if the SD model desires more labor. The vacancy begin rate is computed by the adjusted amounts of labor and vacancy in total (see Eq. (12)). And the adjusted amounts of labor and vacancy are calculated via exponential moving average (smoothing) in Eq. (13) and Eq. (14). Finally, the desired amounts of labor and vacancy are calculated in Eq. (15) and Eq. (16), which are similar to the calculations of desired production and inventory. The decision variables considered in the production module are vacancy creation time (*VacCT*), average time for layoff labors (*LayoffT*), labor fulfillment time (*LaborFT*), and WIP fulfillment time (*WIPFT*).

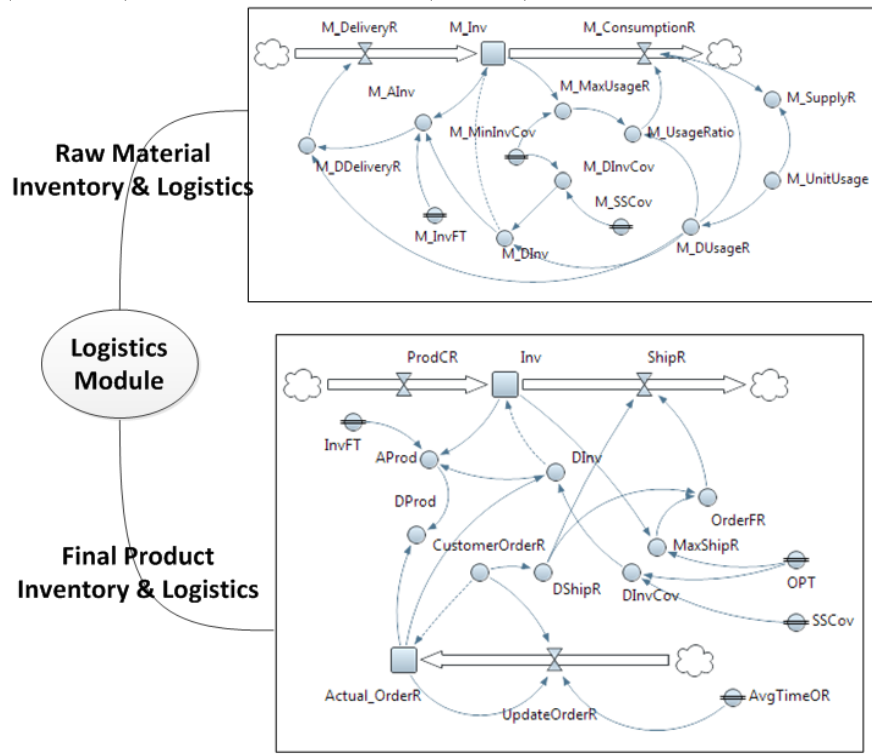

**Fig. 4** Logistics module in the simulation-based game testbed (customized from Sterman (2000))

The logistics part of the SD model is constituted with transportation and inventory control components. As the transportation lead time is simply used, it is translated into inventory fulfillment time for the ease of analysis and the following discussions in this paragraph focus only on the inventory part. Similar to the pro-duction process, one stock is used to aggregate the entire product inventory, and it is calculated in Eq. (17). A retailer maintains an inventory of finished goods, and fills orders as they arrive from customers. The desired shipment rate is set to be equal to the customer demand, while the actual shipment rate depends on the inventory level of the supply chain system. The customer order rate is calculated in Eq. (18), in which the market share and effects of advertisement and promotion are explicitly considered. The un-fulfilled amount of goods will be accounted into backlog inventory, and is calculated for the backorder cost. The order fulfillment ratio is then calculated based on the percentage of order being fulfilled, which is used to decide the actual shipment rate in Eq. (19). Eq. (20) calculates the desired inventory level, which equals to the sum of minimal order processing time and safety level of stock, multiplied by the customer order rate. The variations are also included in Eq. (20). The inventory coverage represents the time duration that the current inventory level under the current shipment rate can cover the customer order, and is a superior measure of both goods holding cost for the supply chain members and the capability of buyers to receive reliable and timely deliveries. This variable is calculated in Eq. (21), and also used to decide the inventory effects to the product price. The decision variables considered in the logistics module are inventory fulfillment time ($InvFT$), raw material transportation lead time ($M\_LT$), product safety stock coverage ($SSCov$), and raw material inventory coverage ($M\_InvCov$).

$$WIP_{i,t} = \int_0^t (ProdBR_{i,s} - ProdCR_{i,s})ds \tag{1}$$

$$AWIP_{i,t} = \lambda^{(W)}(DWIP_{i,t} - WIP_{i,t-1})/WIPFT_i + (1-\lambda^{(W)})AWIP_{i,t-1} \tag{2}$$

$$AProd_{i,t} = \lambda^{(P)}(DInv_{i,t} - Inv_{i,t-1})/InvFT_i + (1-\lambda^{(P)})AProd_{i,t-1} \tag{3}$$

$$DWIP_{i,t} = (AProd_{i,t} + OrderR_{i,t}) \times E(CycleT_i) + \sigma^{(W)} \tag{4}$$

$$DProdBR_{i,t} = AWIP_{i,t} + AProd_{i,t} + OrderR_{i,t} + \sigma^{(P)} \tag{5}$$

$$ProdBR_{i,t} = \max(0, \min(Labor_{i,t} \times ALP_i \times ALT_i, MSR_{i,t}, DProdBR_{i,t})) \tag{6}$$

$$Labor_{i,t} = \int_0^t (HireR_{i,s} - RetireR_{i,s} - layoffR_{i,s})ds \tag{7}$$

$$Vac_{i,t} = \int_0^t (VacBR_{i,s} - HireR_{i,s})ds \tag{8}$$

$$HireR_{i,t} = Vac_{i,t}/E(VacFT_i) \tag{9}$$

$$RetireR_{i,t} = Labor_{i,t}/E(EmployT_i) \tag{10}$$

$$LayoffR_{i,t} = \min(\max(0, -ALabor_{i,t}), Labor_{i,t}/E(LayoffT_i)) \tag{11}$$

$$VacBR_{i,t} = \max(0, ALabor_{i,t} + AVac_{i,t}) \tag{12}$$

$$ALabor_{i,t} = \lambda^{(L)}(DLabor_{i,t} - Labor_{i,t-1})/LaborFT_i + (1-\lambda^{(L)})ALabor_{i,t-1} \tag{13}$$

$$AVac_{i,t} = \lambda^{(V)}(DVac_{i,t} - Vac_{i,t-1})/VacFT_i + (1-\lambda^{(V)})AVac_{i,t-1} \tag{14}$$

$$DLabor_{i,t} = DProdBR_{i,t}/(ALP_i \times ALT_i) \tag{15}$$

$$DVac_{i,t} = \max(0, VacFT_{i,t} \times ALabor_{i,t}) \tag{16}$$

$$Inv_{i,t} = \int_0^t (ProdCR_{i,s} - ShipR_{i,s})ds \tag{17}$$

$$OrderR_{i,t} = TOR \times MS_{i,t} + \sigma^{(O)} \tag{18}$$

$$ShipR_{i,t} = OrderR_{i,t} \times f(Inv_{i,t} / DInv_{i,t}) \tag{19}$$

$$DInv_{i,t} = (OPT_i + SSCov_i) \times (OrderR_{i,t}) + \sigma^{(I)} \tag{20}$$

$$InvCov_i = Inv_i / ShipR_i \tag{21}$$

$$Price_i = E(MP) \times (F_i^{(C)}) \times (F_i^{(I)}) \tag{22}$$

$$F_i^{(C)} = 1 + PSens_i^{(C)} \times (E(c_i^{(P)}) / E(MP) - 1) \tag{23}$$

$$F_i^{(I)} = (InvCov_i / MaxInvCov_i)^{(PSens_i^{(I)})} \tag{24}$$

$$PriceCR = ((Price_1 + Price_2)/2 - MP)/MPFT \tag{25}$$

**Table 1** Nomenclature for system dynamics model

| Notation | Explanation | Notation | Explanation |
|---|---|---|---|
| ProdBR | Production begin rate | SSCov | Safety stock coverage |
| Labor | Labor amount | OrderR | Order rate |
| ALT | Average labor working time per time period | HireR | Labor hiring rate |
| ALP | Average labor productivity per time period | RetireR | Labor retire rate |
| DProdBR | Desired production begin rate | LayoffR | Labor layoff rate |
| AWIP | Adjustment amount for work-in-process (WIP) | VacBR | Vacancy begin rate |
| DProd | Desired production | DHireR | Desired labor hiring rate |
| DWIP | Desired amount of work-in-process | LaborFT | Labor fulfillment time |
| WIP | Amount of work-in-process product | DVac | Desired amount of vacancy |
| *WIPFT* | Fulfillment time for work-in-process product | *AVac* | Adjustment amount for vacancy |
| *CycleT* | Manufacturing cycle time | *MSR* | Raw material supplying rate |
| *ProdCR* | Production complete rate | *Vac* | Labor vacancies |
| *AProd* | Adjustment amount for production | *VacFT* | Average time to fill vacancies |
| *DInv* | Desired inventory level | *EmployT* | Average time of employment |
| *Inv* | Actual inventory level | *MaxLR* | Maximum layoff rate |
| *InvFT* | Fulfillment time for inventory | *VacCR* | Vacancy Closure Rate |
| *OPT* | Order processing time | *DLabor* | Desired labor |
| *InvCov* | Inventory coverage | *ALabor* | Adjustment number of labor |
| *ShipR* | Shipment rate | *LayoffT* | Average time for layoff labors |
| *MS* | Market share | *TOR* | Total order rate |
| *MaxInvCov* | Capacity of inventory coverage | $c^{(P)}$ | Unit production cost |
| *Price* | Product price | *MP* | Market expected price |
| $PSens^{(C)}$ | Price sensitivity to cost | $PSens^{(I)}$ | Price sensitivity to inventory coverage |
| $F^{(C)}$ | Effect of inventory coverage on price | $F^{(I)}$ | Effect of cost on price |
| *PriceCR* | Price changing rate | *MPFT* | Fulfillment time of market expected price |
| $\sigma^{(W)}$ | Variations for desired WIP | $\lambda^{(W)}$ | Exponential smoothing factor for adjusted WIP |
| $\sigma^{(P)}$ | Variations for desired production begin rate | $\lambda^{(P)}$ | Exponential smoothing factor for adjusted production |
| $\sigma^{(O)}$ | Variations for order rate | $\lambda^{(L)}$ | Exponential smoothing factor for adjusted labor |
| $\sigma^{(I)}$ | Variations for desired inventory level | $\lambda^{(V)}$ | Exponential smoothing factor for adjusted vacancy |

\* Subscripts *i* and *t* are omitted. *i* is player index (*i*=A,B), *t* represents simulation replication time.

The product price is determined by Eq. (22) according to Sterman (2000), in which three major parts take effects:

- effect of production costs on price;
- effect of inventory coverage on price;
- impact of retailer/market expected price.

Figure 5(a) depicts the price determination module in the SD model. Inside the price determination mechanism, the effects of duopoly company competition (an enhancement to the original model) is incorporated into the calculation of the retailer expected price. The effect of production costs on price captures the retailer's beliefs on the production costs relative to the expected product price (see Eq. (23)). Either the production cost information ($PSens^{(C)} = 0$) or the retailer's belief ($PSens^{(C)} = 1$) can be ignored depending on the values of sensitivity of price to costs. The effect of inventory coverage on price measures how the relative inventory coverage of supply chain members affects the product price. The sensitivity of price to inventory coverage serves as the exponent of the relative inventory coverage (see Eq. (24)), and its value is negative to reflect the relationship between inventory coverage and price (lower inventory coverage results in higher price). These two equations (Eq. (23) and Eq. (24)) confirm to the original model in Sterman (2000). The third part of the price determination is related with the retailer/market expected price. For a particular type of product, retailers and the consumer market always maintain the belief about the expected price, mainly relying on the past price of similar product. For the simplicity concern, the price biddings among retailer, wholesaler and manufacturer are not explicitly modeled; however, to reflect the price adjustment process over time, the changing rate of market expected price is calculated by the difference of product average price and market/retailer expected price divided by a pre-defined fixed time length (see Eq. (25)). In the price determination process, the experimental control variables considered are price sensitivity to production cost ($PSens^{(C)} \in [0,1]$), price sensitivity to inventory coverage ($PSens^{(I)} \in [-1,0]$), and manufacturer expected price ($Mfg\_Price$).

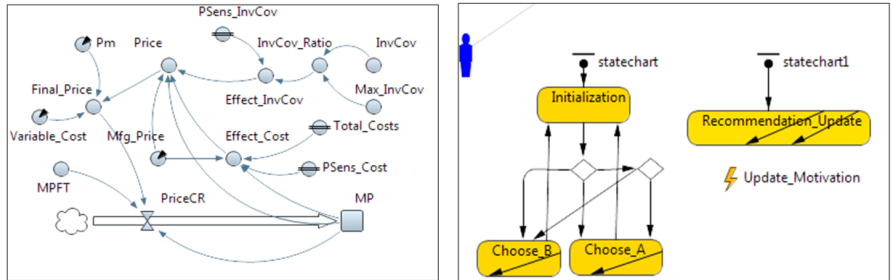

**Fig. 5** (a) price determination module in SD model (left); (b) consumer purchasing behavior in ABS model (right)

## 3.2 Agent-Based Simulation for Modeling Consumer Purchasing Behavior

Figure 5(b) is the module snapshot of the consumer purchasing behavior in ABS model. Equations from (26)-(36) represent underlying mathematical relationships of the module, and Table 2 provides nomenclatures for variables and parameters used in those equations. For the marketing expense, it is assumed to have two aspects: advertisement and promotion. Eq. (26) and Eq. (27) are used to calculate the spending for advertisement and promotion over a considered time period (i.e. a period for a certain marketing strategy). The amount of marketing budget is decided according to the company's revenue. The Chief Marketing Officer (CMO) Council report (2010) demonstrates a direct relation between marketing budget and revenue for various companies. Based on our case study (i.e. soft drink duopoly), the corresponding percentage of marketing investment is selected. An adjustment factor is introduced in these two equations to ensure that a realistic scenario (e.g. order of magnitude) can be achieved. The market spending rate is then derived (see Eq. (28)) by incorporating the adjustment time for spending market budget into the calculation. The decision variables considered in the marketing strategy are marketing budget ($MB$), advertisement intensity ($Ad$), and promotion depth ($Pm$).

Marketing force concept in this work has been adopted from extended Lanchester model (Naik *et al.*, 2005) and is depicted in Eq. (29). The marketing force depends not only on the weight of advertisement intensity, promotion efforts (e.g. frequency and depth), but also on their marketing strategy interactions that are discussed in details in the next paragraph. The marketing force is the leading power, which influences the consumer's perception (e.g. sensitivity of promotion, susceptibility of advertisement) of a particular product. In this chapter, these relationships are captured in Eq. (30), Eq. (31), and Eq. (32). As the market force is dynamically updated through the simulation run, the consumer's perception is also updated according to the change of market force. This assumption implies that if a company loses most of the market, it would have to sacrifice even more to win back the market share.

Another important feature in the ABS model is that we explicitly incorporate the marketing interaction effects between companies. These marketing interactions include the binding constraints on the sum of expenditures on the advertisements and promotions, as well as the segregation of locations and communication channels expressed in terms of expenses (Naik *et al.*, 2005). In this work, the mathematical formulation is based on these concepts, where the interactions for each pair of activities are explicitly modeled. To take the strategic foresight of manager into account, co-state dynamics in Eq. (33) is adopted, and the interaction effects between companies are formulated as the co-state variables. The values of co-state variables in the next time point are captured by the differential equation given the current interaction effects. Then, the sunk cost is calculated (see Eq. (34)), which incurs due to the strategic interactions between duopoly companies. The case study presented in Sect. 5.1 provides more details on these interactions in the context of a soft drink duopoly competition.

In a consumer market, consumers make the adoption decision based on various factors from both the companies and environment, such as unit price, advertisement, promotion, quality, and word-of-mouth recommendations. In our simulation model, it is assumed that an agent (i.e. consumer) becomes an adopter of a particular product based on the motivation function incorporating effects of four factors—price sensitivity, advertisement influence, promotion sensitivity, and acquaintance influence. Based on the model in Zhang and Zhang (2007), we proposed an improved formula to calculate the consumer motivation to purchase brand $i$ at time point $t$, in which the motivation value is decided by the following three attributes of price, advertisement intensity, and agent influence exerted by other agents (consumers). The enhancements made in this work are as follows:

- incorporation of a social network structure to represent interactions among agents;
- incorporation of advertisement and promotion factors to mimic more realistic decision making process.

In our study, a scale-free social network model called Barabasi-Albert Model (BA model, also known as Preferential Attachment Model) (Albert and Barabasi 2002) is built to represent the social relationships of customers for the artificial market. The BA model reflects the "rich-get-richer" phenomenon in societies and the degree of nodes follows a power-law distribution, in which the probability of a new node connecting to an existing node is proportional to the degree of it. To incorporate the advertisement and promotion effects from marketing activities into the consumer purchasing decision, the price sensitivity, susceptibility to advertisement, promotion-sensitivity and follower tendency, have been set to associate with price, advertisement, promotion, and recommendation influence, respectively. The initial value of susceptibility to advertisement, promotion-sensitivity and follower tendency are pre-set at the beginning of simulation run. The price sensitivity is an exponential function of the difference between the real price of a product and the expected average price of the product (see Eq. (35)). In this equation, $s$ is a price parameter ($s>1$), and takes the same values for the similar competitive types of product, $m$ is a constant and its value is based on an agent's socio-economic attributes (e.g. millionaires are less price sensitive than unemployed persons). The consumer purchasing motivation function is calculated in Eq. (36), which decides the product selection of consumers. It is assumed that agents will always select a product having a higher motivation value, and randomly choose one if the motivation values are equal.

$$AdS_i = K \times MB_i \times Ad_i \qquad (26)$$

$$PmS_i = K \times MB_i \times Pm_i \qquad (27)$$

$$MSR_i = (AdS_i + PmS_i)/AdjTimeMS \qquad (28)$$

$$MF_i = \omega_1 Ad_i + \omega_2 Pm_i + \omega_3 Ad_i Pm_i + Inter_i \tag{29}$$

$$SusAd_i = MF_i \times I_a \tag{30}$$

$$SensPm_i = MF_i \times I_p \tag{31}$$

$$Ft_i = MF_i \times I_f \tag{32}$$

$$\begin{pmatrix} \overline{inter_1} \\ \overline{inter_2} \end{pmatrix} = \begin{pmatrix} \delta_1 & \delta_1\delta_2 \\ \delta_2\delta_1 & \delta_2 \end{pmatrix} \begin{pmatrix} \rho+F & 0 \\ 0 & \rho+F \end{pmatrix} \begin{pmatrix} inter_1 \\ inter_2 \end{pmatrix} - \begin{pmatrix} Price_1(1-Pm_1) \\ Price_2(1-Pm_2) \end{pmatrix} \tag{33}$$

$$C^{(s)} = \sum_{i=1}^{2} MB_i \times Inter_i \tag{34}$$

$$SensP_i = -s^{Price_i \times (1-Pm_i) - (Price_1 + Price_2)} + m \tag{35}$$

$$M_i = SensP_i \times Price_i(1-Pm_i) + SusAd_i \times Ad_i + SensPm_i \times Pm_i + Ft_i \times Inf_i \tag{36}$$

**Table 2** Nomenclature for agent-based model

| Notation | Explanation | Notation | Explanation |
|---|---|---|---|
| $MB$ | Marketing budget | $\rho$ | Co-state parameter |
| $Ad$ | Advertisement intensity | $\delta$ | Co-state factor |
| $Pm$ | Promotion depth | $C^{(s)}$ | Marketing sunk cost of the duopoly companies |
| $K$ | Adjustment factor | $Inf$ | Follower tendency influence |
| $AdS$ | Spending rate on advertisement | $M$ | Customer purchasing motivation function |
| $PmS$ | Promotion spending rate | $SensP$ | Price sensitivity |
| $AdjTimeMS$ | Adjustment time for marking budget spending | $SensPm$ | Sensitivity of consumer to promotion |
| MSR | Marketing spending rate | SusAd | Susceptibility of consumer to the advertisement |
| $\omega_i, i=1,2,3$ | Weights of market force effects | Ft | Follower tendency |
| Inter | Interaction effect between two duopoly companies | s, m | Price sensitivity parameters |
| MF | Marketing force | $I_a, I_p, I_f$ | Initial value of SusAd, SensPm, and Ft. |
| F | Total marketing force | Mfg_Price | Manufacturer expected price |

\* Subscripts $i$ and $t$ are omitted. $i$ is player index ($i$=1,2), $t$ represents simulation time.

## 3.3 Payoff in Simulation-based Duopoly Game

The total net profit serves as the payoff of simulation-based game, which is calculated in Eq. (37). The cost items constituting the payoff function based on the simulation outputs are depicted in Table 3. All different cost items across the production, logistics and marketing activities are considered in the payoff function, and the time length to calculate all the cost items is the total simulation replication length. After the simulation run, the outputs are collected to calculate the net profit earned for each company. A payoff matrix is then constructed based on the outputs and is used to approximate the best response (i.e. equilibrium) of the duopoly game, which will be discussed in Sect. 4.

$$Payoff_i = TRev_i - (C_i^{(P)} + C_i^{(R)} + C_i^{(I)} + C_i^{(B)} + C_i^{(T)} + C_i^{(M)} + C^{(S)}) \tag{37}$$

Table 3 Nomenclature for game payoff components

| Payoff components | Descriptions |
|---|---|
| $TRev_i$ | Total revenue for product $i$ |
| $C_i^{(P)}$ | Total production cost for product $i$ |
| $C_i^{(R)}$ | Total raw material purchasing cost for product $i$ |
| $C_i^{(I)}$ | Total inventory cost for product $i$ |
| $C_i^{(B)}$ | Total backlog cost for product $i$ |
| $C_i^{(T)}$ | Total transportation cost for product $i$ |
| $C_i^{(M)}$ | Total marketing spending for product $i$ |

## 4. Simulation-based Game Solving and Analysis

In this section, a detailed simulation-based game solving and analysis (GSA) procedure proposed in this chapter will be discussed. The intent of the proposed procedure is to make the problem tractable by restricting the profile strategies that each company is allowed to play without losing the generalization from the original game. Large strategy spaces consist of continuous and multi-dimensional action sets, while the perfect information assumption is hold to reduce the problem complexity for analysis. Due to the symmetric property of the game, two agents are assumed to have identical behavior possibilities, and be exposed to the same customer market. Before discussing the details of the GSA components and procedure, notations regarding a normal form game, simulation-based game and the equilibrium concepts are introduced first.

### 4.1 Setup and Motivation

A *normal form game* can be formally expressed as $\Gamma = [I, \{s_i, \Delta(s_i)\}, \{u_i(s)\}]$, where $I$ refers to the set of players and $I = \{1, 2\}$ in our study; $s_i$ and $\Delta(s_i)$ denotes the pure and mixed strategy for player $i$ ($i \in I$) respectively; $u_i(s)$ is the payoff function of player $i$ when strategy profile $s$ has been selected. An important variable frequently used in analyzing normal form game is *regret* of a profile $s \in S$, denoted by $r(s)$, which is calculated in Eq. (38).

$$r(s) = \max_i \max_{s_i'} u_i(s_i', s_{-i}) - u_i(s) \tag{38}$$

In Eq. (38), $s_i' \in \{S_i - \{s_i\}\}$ and $s_{-i}$ represents for a strategy profile other than that of player $i$. Next, definition regarding game solution is given as follows: a *Nash Equilibrium* of the normal-form game is a strategy profile $s \in S$ such that for every player $i \in I$, Eq. (39) holds.

$$u_i(s) \geq u_i(s_i', s_{-i}), \forall s_i' \in S_i \tag{39}$$

In this chapter, Nash equilibrium, equilibrium, and game solution terms are used interchangeably. Furthermore, the *symmetric game* setting is also considered, in which the following two conditions need to be satisfied:

- $S_i = S_j$ for all players $i, j \in I$;
- $u_i(s_i, s_{-i}) = u_j(s_j, s_{-j})$ for every $s_i = s_j$ and $s_{-i} = s_{-j}$.

In addition, the terms of simulation-based game and empirical game are used interchangeably because they essentially convey identical meanings. A *simulation-based game* is defined that the player's payoff is specified via simulation models, and the definition of *empirical game* is focused on estimating the payoff matrix using simulation outputs (Vorobeychik, 2008). In the empirical game setting with a large number of strategy profile and noisy samples involved, calculating the exact Nash Equilibrium is sometimes intractable. Another way of approximating it is applying $\varepsilon$-Nash Equilibrium ($\varepsilon$: tolerance), which is a profile $s \in S$ satisfying Eq. (40) for every player $i \in I$.

$$u_i(s) + \varepsilon \geq u_i(s'_i, s_{-i}), \forall s'_i \in S_i \tag{40}$$

As the game is constructed in simulation, we differentiate two types of payoff: the *true payoff* existing in a real practice duopoly and the *estimated payoff* obtained from simulation outputs. When constructing an empirical payoff matrix, a simulation model will be run to obtain noisy samples for each pure or mixed strategy profile. The noisiness in the sample includes the randomness from the simulation experiments as well as the players' mixed strategies (Vorobeychik, 2010). Empirical game is the one, which maintains the same strategy profiles for all players while the payoffs of them involve noise. For each specific profile of any player in an empirical game, the payoff is an estimate value by taking arithmetic mean of multiple data points from the noisy sample as shown in Eq. (41).

$$\hat{u}_{i,n}(s) = \sum_{j=1}^{n} U_{ij}(s) \bigg/ n \tag{41}$$

The equation shows an estimate of payoff to player *i* for profile strategy *s* based on *n* samples. From now on for the terminologies used in our discussions, readers are suggested to refer to Table 4.

**Table 4** Clarification of terminologies used

| Name | Definition | Explanation |
| --- | --- | --- |
| Aggregated strategic factor | The factor including aggregated information of other factors | *Production* factor, *Logistics* factor, |
| Detailed strategic factor | The factor decomposed from aggregated strategic factor | *Order lead time*, *safety stock coverage* decomposed from logistics factor |
| Strategic factor levels | The different levels (i.e. values) that a factor can achieve/attain | (H) for high level of production factor |
| Strategy | Combination of different levels for a group of strategic factors* | (H, L, L, H)* |
| Profile | Combination of strategies chosen by game players | {(L, L, H, H)$_1$, (H, L, H, L)$_2$} is one profile for a two-player game |
| Solution profile | Players' profile obtained when game reaches the equilibrium | Element(s) in the profile set |
| Solution payoff | Players' payoff obtained when game reaches the equilibrium | Element(s) in the payoff matrix |
| True payoff | The ideal payoff for the game player | $u(s)$ with respect to profile $s$ |
| Estimated payoff | The estimated payoff value obtained from simulation | $\hat{u}_{i,n}(s)$ with respect to profile $s$ by running $n$ simulation samples |

*$H$: high, $L$: low; strategic factor: decision variable.

The academic challenge of solving such a game is that the constrained simulation and experimental resource cannot afford the enormous number of strategies and samples. According to the discussion in Sect. 3, the duopoly game includes totally 12 strategic factors for each player: if every single strategic factor takes only two levels, the total number of profiles in the entire profile set under symmetric game setting is $(2^{12})^2/2=8,388,608$. Assuming each simulation replication takes 1 second and only 10 replications are taken for each individual profile, the total time needed to complete the simulation of the entire profile set would be 2.66 years, which is unrealistic to perform in practice.

The above computational challenge motivates development of the GSA procedure in this chapter. As it is impractical to construct a comprehensive payoff matrix and achieve the exact game equilibrium(s) by involving all strategic factors (and their levels), targeting on the critical factors that can approximate the true equilibrium becomes the major undertaking. As the number of profiles is reduced, the sample size for each profile can be increased accordingly. The trade-offs between *strategy refinement* and *data sampling* is: given a fixed amount of simulation/experimental resources, exploring more profiles decreases the number of samples that can be chosen, which may influence the accuracy of estimated game payoff by the end; while more samples will restrict the span of profiles to be selected, which may rule out the key strategies that will impact the game solution eventually. Other than the strategy refinement and data sampling, a game solving engine/algorithm and performance evaluation criteria are also needed to complete the GSA.

### 4.2 Simulation-based Game Solving and Analysis

To resolve the formulated simulation-based game, four components are required:

- First of all, an approach to explore and refine the strategy space is needed. As discussed before, some strategies are more significant to determine the game equilibrium than others. Our objective here is to explore those critical strate-gies in a more detailed manner so that insights can be gained on how the key strategic factors can impact on the game equilibrium.
- Second, a method to decide the sampling procedure is needed. As known, sampling cost and information gain are always the trade-offs during the sampling procedure. Given the sampling resource availability and capability, the sampling procedure should be able to achieve the maximum information gain so as to better approximate the true game payoff.
- Third, a game solving engine is needed, which will allow us to find equilibrium(s) for the simulation-based game under different initial game settings (e.g. initial strategy profile, problem scenario). The game solution should include pure, mixed or both types of equilibriums.
- Forth, evaluation criteria for assessing the performance of GSA procedure is needed, which will capture the main features of the GSA procedure by dealing with the game equilibrium results. The evaluation criteria should also contain the assessments of major equilibrium properties (e.g. weakness, strictness, stability, and robustness).

We first formulate an algorithm, which depicts how these four components mentioned above work together to solve and analyze the simulation-based game. Then detailed contents on each component are discussed. Note that each round of the GSA procedure run is called an *iteration*. The GSA procedure includes the following major steps:

**Step 1** Develop an initial game *profile set* by selecting strategic factor levels, then choose an initial *sample size* for each profile and set $g$ equals to 1.
**Step 2** Run the simulation model based on the selected profile set and sample size, construct the *empirical payoff matrix* according to the simulation outputs.
**Step 3 (Game Solving)** Solve the game for pure strategy equilibrium by improving the *unilateral deviation set* for each player one after the other until no more improvements can be obtained.
**Step 4 (Strategy Refinement)** Employ *design of experiments* technique to decide the statistical significance of each *aggregated strategic factor* with respect to the game payoff. Then, if $g$ equals to 1, go to **Step 4.1**; if $g$ equals to 2, go to **Step 4.2**.
  **Step 4.1** Include all the *detailed strategic factors*, which are decomposed from the current *significant* aggregated strategic factor, into the *refined profile set*; eliminate the insignificant aggregated strategic factor(s) from the *refined profile set*. If no more detailed strategic factors can be included, go to **Step 5** and set $g$ equals to 2.
  **Step 4.2** Include more strategic factor levels into the *refined profile set* for the next *iteration*, go to **Step 5**. If no more levels for each strategic factor need to be added, terminate the GSA process and go to **Step 6**.

**Step 5 (Data Sampling)** Given the significant strategic factors and their levels in the *refined profile set*, decide the sample size for each profile using the *enhanced ECVI sampling* approach. Go to **Step 2**.

**Step 6 (Performance Evaluation)** Based on the game equilibrium results, calculate values for all the *evaluation criteria* inside and between GSA iterations, and summarize the results.

Steps 1 and 2 are mainly for algorithm initialization and payoff generation, respectively. An indicator variable $g$ is used in Step 4, which represents the refinements of either strategic factors ($g=1$) or factor levels ($g=2$). Provided that a reasonable experimental time and cost can be spent on the simulation experiments, the trade-offs between the strategy refinement extent and data sampling size always exist. Table 5 provides comparison results with varying numbers of strategy refinement and sampling size given a fixed affordable experimental time (i.e. 5 days) for the simulation run. The lower limit of the experimental cost is bounded by ensuring a minimum degree of strategy refinement and sampling size, while the upper limit is related with the total affordable experimental cost. As shown in Table 5, if each experimental iteration is selected to be 5 days, a total of four strategic factors can be selected to ensure a reasonable number of samples (i.e. 150) in the experiments.

**Table 5** Trade-offs between strategy refinement and data sampling

| Total strategic factors | No. of strategies for each player (level = 2) | No. of profiles to be evaluated | Affordable experimental time limit (days) | Time per simulation replication (seconds) | No. of samples affordable for each profile |
|---|---|---|---|---|---|
| 1 | 2 | 3 | 5 | 20 | 7200 |
| 2 | 4 | 10 | 5 | 20 | 2160 |
| 3 | 8 | 36 | 5 | 20 | 600 |
| 4 | 16 | 136 | 5 | 20 | 158.82 |
| 5 | 32 | 528 | 5 | 20 | 40.90 |
| 6 | 64 | 2080 | 5 | 20 | 10.38 |
| 7 | 128 | 8256 | 5 | 20 | 2.62 |
| 8 | 256 | 32896 | 5 | 20 | 0.66 |
| 9 | 512 | 131328 | 5 | 20 | 0.16 |
| 10 | 1024 | 524800 | 5 | 20 | 0.04 |

The strategy refinement method essentially seeks to find out in which order and with what specific strategic factor levels to include the strategies to the simulation-based game analysis. It is slightly different to the strategy exploration problem in Jordan *et al.* (2008), with the modification of the word "refinement" that is tightly related with both the game strategy and simulation modeling details. As noted before, each strategic factor (e.g. production) involves different detailed aspects (e.g.

labor control, raw material procurement). The strategic factors that are more significant than others should be considered with priority in the simulation testbed and also decomposed into more detailed levels for analysis. The purpose of doing so is to approximate the game equilibrium without evaluating all the strategy profiles, which is time-consuming, cost-inefficient, and even intractable. The strategy refinement process, which starts from an aggregated level and then moves to a more detailed level, is set as follows:

- For the initial experiment, the focus of the profile set (simulation inputs) is only at the aggregated strategic factors (e.g. production, logistics), and multiple (e.g. 2) levels of these factors are selected for experimental study.
- *Design of experiments* technique is then used to identify the critical strategy profiles by analyzing the simulation outputs. Figure 6 depicts the process, in which the inputs to the experimental design is the different levels of strategic factors and the empirical payoff matrix generated from simulation outputs, while the outputs of the experimental design are the factors that have significant impacts on the game payoff.
- Then, for those critical strategies, more insights on how different values of strategic factors impact the game payoff are investigated via partitioning the factors into detailed factors or levels depending on the requirements. Then, we treat each strategic factor or level as the input to the simulation for the next experiment iteration.

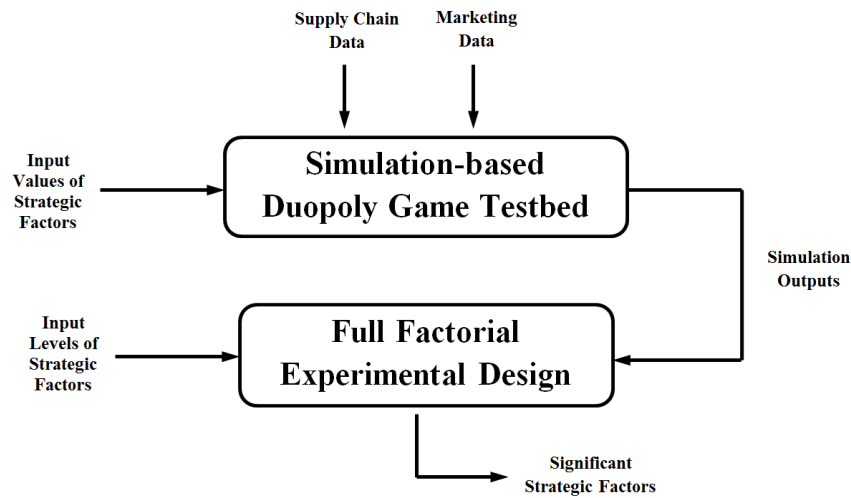

**Fig. 6** Experimental design for strategic factor refinement via simulation

The above mentioned process (i.e. empirical payoff generation via game simulator, identification of significant factors via experimental design) is applied iteratively in GSA procedure. During the iterative process, game is solved and the immediate results are used to find the corresponding profiles for sampling. It is noted that under different simulation scenarios, the outputs of experimental design may be different. In addition, various experimental design techniques may be applied as long as they provide better insights into the analysis. This work employs a standard two-level full factorial experimental design technique as a pilot study for strategy refinement.

For sampling significant profiles, an approach named *estimated conformational value of information* (ECVI) in Walsh *et al.* (2003) has been enhanced in this chapter. The ECVI measures the degree to which further samples would reduce the estimated error (denoted by $\hat{e}(x)$) of the current equilibrium solution (see Eq. (42) and Eq. (43)).

$$\hat{e}_{i,p}(x(s)) = \hat{u}_{i,p}(s) - u_i(s) \tag{42}$$

$$ECVI(q \mid i, s, p) = E_{q\mid p}[\hat{e}_{i,p}(x(s)) - \hat{e}_{i,p.q}(x(s))] \tag{43}$$

In Eq. (42) and Eq. (43), $s$ represents a strategy chosen to conduct sampling, $p$ and $q$ refer to the number of data points being sampled and to be sampled, respectively. The maximum information gain is achieved by selecting the maximum value of ECVI, which also indicates the best choice of samples. This method has been chosen in our study as it has been approved to show significant improvement over the uniform sampling method. While the criteria for stopping sampling and the tradeoff between the sampling cost and information value gain are not discussed in details in Walsh *et al.* (2003), our work addresses them explicitly. The sampling cost mainly depends on the simulation replication length, and the information value gain refers to how important more samples can help to make an accurate decision. In this chapter, two separate items in ECVA are classified in the GSA procedure:

- A pre-selected threshold value of affordable sampling size, which is the maximum number of samples that can afford to run for each profile based on the experimental resource availability.
- A lower limit of *ECVI gain*, which is designed by user and aimed to ensure the game solution quality.

Under the two criteria discussed above, we want to find the corresponding sample size either satisfying the lower limit of ECVI gain (denoted by $ECVI_s^{(L)}$), or reaching the limit of sampling capability (denoted by $N_s$), as shown in Eq. (44).

$$p + q = \min(N_s, ECVI_s^{(L)}) \tag{44}$$

This enhancement provides flexibility to users, where they can select their own threshold values depending on the experiment requirements. In our experimental study, we have applied this approach to eliminate the twisted sample values (the extreme low and high values), which constitute about 10% of all data samples.

Integration of strategy refinement and data sampling discussed so far in this section contributes to the uniqueness of the proposed GSA approach. This integration allows us to combine the advantages of both, as well as to avoid the potential drawbacks of spending additional simulation resources for sampling all profiles. The next step in our procedure is to input the selected game strategy profile and sample size into the simulation-based game testbed. The simulation outputs are then collected to construct the empirical payoff matrix. Then, we apply a game solving engine to calculate the pure Nash Equilibrium for the duopoly players. The game solving engine computes the equilibrium by improving the *unilateral deviation* set in Eq. (45) for each player one after another, until no more payoff gain can be obtained.

$$D_i(s) = \{(\breve{s}_i, s_{-i}) : \breve{s}_i \in S_i\}, i = 1, 2 \tag{45}$$

This is a traditional approach, but still the most effective and efficient way to obtain the pure Nash Equilibrium. As the empirical payoff matrix always involves variations, an $\varepsilon$-Nash Equilibrium concept (see Eq. (40)) is used to ensure that the potential optimum solutions are included during each experiment iteration.

As the game solution involves variations due to different reasons such as limited simulation/experimental resources and sampling errors inherent to simulation, proper criteria on assessing the GSA procedure has been developed in this chapter. As mentioned earlier, the GSA procedure stops when no more iteration (e.g. strategy refinements) can be established. As each experiment iteration proceeds and the simulation gains more fidelity (details), we intend to find 1) whether the equilibrium stays unchanged or evolves to be better (e.g. strictness vs. weakness), 2) how the modeling details can impact the game payoff, and 3) how sensitive the equilibrium(s) are to the disturbances. The evaluation criteria developed in this work focus on the following aspects:

- *confidence intervals* of the game equilibrium(s) for examining the closeness of estimated and true payoffs (See Eq. (46));

$$\Pr(\hat{u}_{i,n}(s) - \theta < u_i(s) < \hat{u}_{i,n}(s) + \theta) = 1 - \alpha \tag{46}$$

- *statistical test* (i.e. two-sample *t*-test) for evaluating the differences between solution profile and its neighboring profiles;
- *statistical test* (i.e. two-sample *t*-test) for evaluating the differences between solution payoffs over iterations;

- experimental studies on the stability of the game equilibrium(s): the equilibrium stability concepts applied here are originated from Szidarovszky and Bahill (1998), and we define three types of stability as follows:

  – *Asymptotic stability* with respect to game equilibrium refers to that for a given initial game state (i.e. players' initial profile), the player payoff for the solution profile eventually converges to the solution payoff.
  – *Marginal stability* with respect to game equilibrium is the one that for a given initial game state, the player payoff for the solution profile converges to a region containing the considered solution payoff and its tolerance.
  – *Instability* with respect to game equilibrium refers to the players' profile that does not belong to the above two categories.

## 5. Experiments and Results

### 5.1 Soft-drink Duopoly Experiment Setup

Under the current market scanner, the soft drink industry exhibits a classic example of duopolies involving integrated supply chain and marketing activities. Cola wars between The Coca-Cola Company® and PepsiCo Inc® and related literature works (Morris, 1987) have served as a basis for our case study. The two companies together account for about three-quarters of the total soft drink market share. In fact, the industry has high operational overlap since different suppliers and manufacturers (e.g. producers and bottlers) possess similar impetus of sales and profits along the supply chain, and in the market side a similar customer base is shared for the duopoly companies. While the soft drink industry as a whole enjoys positive economic profits among all of its members, the ultimate goal for the industry should be to create a win-win situation for both the manufacturers as well as the customers.

As mentioned earlier, both Coca-Cola Company® and PepsiCo Inc® mainly trade on supply chain and marketing values, and invest substantial portion of their revenues in those areas. Modeling of the major activities in those areas has been discussed in Sect. 3. Different values of the decision parameters for the proposed simulation model depict the various scenarios encountered in the soft drinks duopoly. Table 6 shows the strategic factor values used in the experiments of this chapter, and the length of simulation replication run is about 3 months (100 days). We then estimate a payoff matrix through the constructed normal-form simulation-based game, with the emphasis on the strategies mentioned in Table 6.

**Table 6** Strategic factors and levels used in experiments

| Aggregated strategic | Detailed strategic factor | Strategic factor levels | |
|---|---|---|---|
| | | Low* | High* |

| factor | | L | ML | MH | H |
|---|---|---|---|---|---|
| Manufacturing | Vacancy creation time (days) | 1 | | 5 | |
| | Average time for layoff labors (days) | 3 | | 7 | |
| | Labor fulfillment time (days) | 4 | | 12 | |
| | WIP fulfillment time (days) | 1 | | 3 | |
| Logistics | Inventory fulfillment time (lead time)(days) | 2 | 6 | 10 | 14 |
| | Raw material transportation lead time (days) | 1 | 3 | 5 | 7 |
| | Product safety stock coverage (days) | 2 | 6 | 10 | 14 |
| | Raw material inventory coverage (days) | 1 | 3 | 5 | 7 |
| Pricing | Price sensitivity to production cost ($\in [0,1]$) | 0.1 | | 0.9 | |
| | Price sensitivity to inventory coverage ($\in [-1,0]$) | -0.1 | | -0.9 | |
| | Manufacturer expected price ($) | 1 | | 2 | |
| Marketing | Marketing budget (MB) ($) | 5% of revenue | | 15% of revenue | |
| | Promotion depth (% of MB, uniform distributed) | (0.1,0.2) | (0.2,0.3) | (0.3,0.4) | (0.4,0.5) |
| | Advertising intensity (% of MB, uniform distributed) | (0.1,0.2) | (0.2,0.3) | (0.3,0.4) | (0.4,0.5) |

*L: low, ML: medium low, MH: medium high, H: high.

Table 7 shows the strategic factors and levels involved in each experiment iteration. To balance the trade-offs between strategy refinement and data sampling, 16 strategies for each player (4 strategic factors) and 70 initial data samples for each profile are selected during each experiment iteration (the total number of profile is 16*16=256). As the considered game is symmetric, only the upper triangular of the strategy matrix is used for sampling, which is equivalent to 136 ([(16*16)-16]/2+16=136) profile sets. After applying the modified ECVI data sampling approach, samples with roughly 10 upper and 10 lower extreme values have been eliminated for each profile. So the effective sample size in our experi-ment is 50. As each iteration may involve different strategic factors (aggregated or detailed), notation $(S_m^{(k)}, S_n^{(k)})$ is used to represent the profile information for player A selecting strategy $m$ ($m = 1, 2, ..., 16$) and player B selecting strategy n ($n = 1, 2, ..., 16$) during $k^{th}$ iteration.

**Table 7** Strategic factors used over GSA iteration in experiments

| Iteration | Strategic factors | Strategic factor levels* | Iteration | Strategic factors | Strategic factor levels* |
|---|---|---|---|---|---|
| 1st | Manufacturing | L/H | 2nd | Advertising intensity | L/H |
| 1st | Logistics | L/H | 3rd | Raw material inventory coverage / Product safety stock coverage | L/ML/MH/H |
| 1st | Pricing | L/H | 3rd | Raw material transportation lead time / Inventory fulfillment time | L/ML/MH/H |
| 1st | Marketing | L/H | 4th | Raw material inventory coverage / Product safety stock coverage | L/ML/MH/H |
| 2nd | Raw material inventory coverage / Product safety stock coverage | L/H | 4th | Promotion depth | L/ML/MH/H |
| 2nd | Raw material transportation lead time / Inventory fulfillment time | L/H | 5th | Raw material inventory coverage / Product safety stock coverage | L/ML/MH/H |
| 2nd | Promotion depth | L/H | 5th | Advertising intensity | L/ML/MH/H |

*L: low, ML: medium low, MH: medium high, H: high.

### 5.2 Experiment Results

In this section, we describe the experimental results and demonstrate the effectiveness of the proposed GSA procedure under the hybrid simulation framework. For the limited space, only pure strategy equilibrium(s) are analyzed in this section.

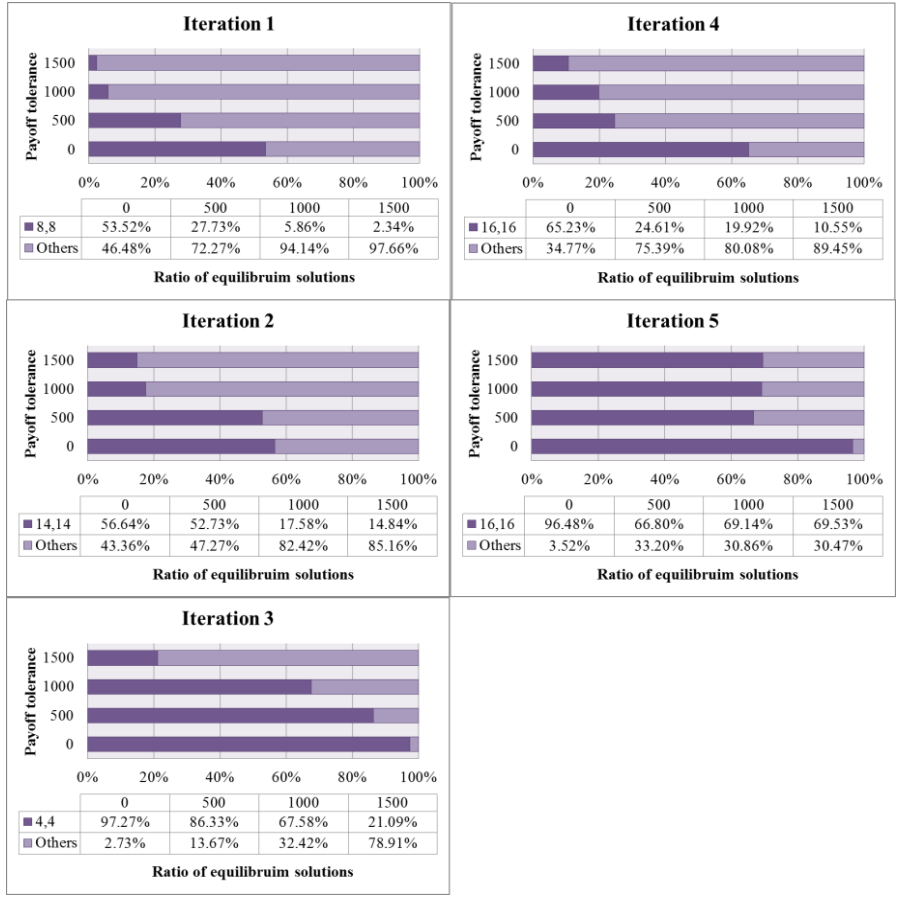

**Fig. 7** Evolution of game equilibriums over GSA iterations

Figure 7 depicts the percentage of game equilibriums computed for all 5 iterations, in which the horizontal axis represents the solution percentage and the vertical axis represents the payoff tolerance. As it is a duopoly game, it is highly believed that the solution profile has the symmetric strategy for the two players (i.e. $(S_{n_1}^{(k)}, S_{n_2}^{(k)})$ with $n_1 = n_2$). That's the reason why we only draw the symmetric strategy in Figure 7, and notify other potential solution strategies as "Others". In Figure 7, we observed that as the payoff tolerance increases within each iteration, the empirical game tends to involve more equilibriums than the case under zero tolerance. As the tolerance value is highly related with the sample size of each profile, it is difficult to reduce the tolerance by sampling more data (sampling cost is limited). However, the suspected solution profile with its neighborhood strategies, which only involves roughly 8 to 12 data points, can be extracted out and sampled with more data points. Another observation from Table 8, which conforms to our intuition, is when the sample size enlarges from 50 to 500, the half width of confidence interval for each potential solution payoff reduced. As the iteration proceeds (from Iteration 1 through 5), the half width of confidence interval (CI) also decrease, which indicates the estimated equilibrium is closer to the true equilibrium. However, under the sample size of 500, the decreasing trend of the CI values is not as salient as that for the case with the sample size 50 or under.

Table 8 Comparisons of solution profiles and payoffs over GSA iterations

| Iteration | $ES_1$ | $ES_2$ | Payoff for player 1 | | | | Payoff for player 2 | | | |
| --- | --- | --- | --- | --- | --- | --- | --- | --- | --- | --- |
| | | | Sample size=500 | | Sample size=50 | | Sample size=500 | | Sample size=50 | |
| | | | Mean | HW | Mean | HW | Mean | HW | Mean | HW |
| 1st | 8 | 8 | 23967 | 449 | 23333 | 1257 | 23922 | 467 | 23265 | 1184 |
| 2nd | 14 | 14 | 24438 | 450 | 23952 | 1129 | 24452 | 461 | 24401 | 1073 |
| 3rd | 4 | 4 | 24854 | 422 | 25829 | 924 | 24845 | 441 | 24539 | 822 |
| 4th | 16 | 16 | 25428 | 408 | 23778 | 915 | 25505 | 415 | 25671 | 841 |
| 5th | 16 | 16 | 26178 | 397 | 26964 | 729 | 26305 | 408 | 28235 | 866 |

*$ES_i$: Equilibrium for player $i$

To evaluate the game equilibrium robustness (weakness vs. strictness), a group of 10 samples close to the estimated Nash Equilibrium (its neighborhood that have the similar payoff values with it) have been selected in each iteration. Two sample t-tests (hypothesis testing $H_0: \mu_1 = \mu_2$, $H_1: \mu_1 \neq \mu_2$) are then performed on each pair of selected data samples, followed by the two-tailed P-value calculation. Figure 8 organizes the calculated P-values into the box-plot, in which a reduced trend of major portion (25%~75%) and the median of data are observed over iterations for both player A (left) and B (right). In other words, initially (Iteration 1 or 2) the game equilibrium is not significantly different from its neighborhood values; while after several iterations, the game equilibrium is almost all significantly different from its neighborhood values (Iteration 5). From the results of box-plots, a conclusion can be made: the game equilibrium(s) evolves from weak to strict during iterations of the GSA procedure.

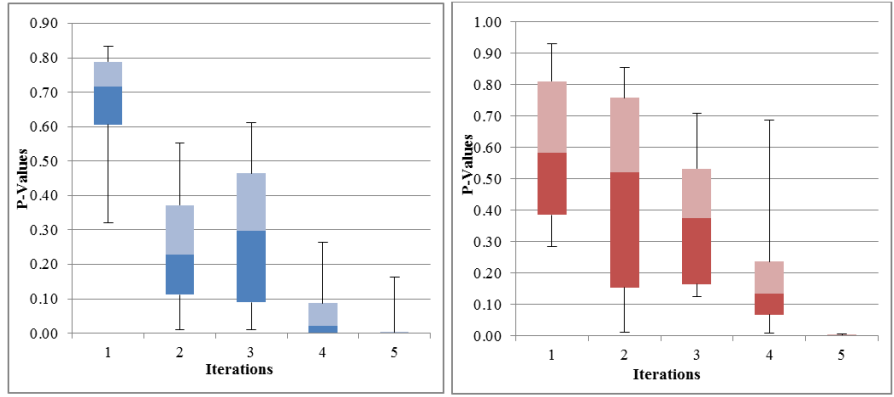

**Fig. 8** Box-plots for the *P*-values of two sample *t*-test on solution profile with its neighbor profiles over GSA iterations: player A (left); player B (right).

Another statistical test involves the equilibrium comparisons over different iterations. As there seems an increasing trend of equilibrium payoff over iterations, this test helps to identify how significantly different each pair of equilibrium payoffs is. The one-sided hypothesis testing is constructed with $H_0 : \mu_1 = \mu_2$, $H_1 : \mu_1 < \mu_2$; and the comparisons are performed between iterations. Figure 9 shows the comparison results in a bar chart, where the horizontal axis numbers (1 through 7) correspond to comparison groups of (1 vs. 2), (2 vs. 3), (3 vs. 4), (4 vs. 5), (3 vs. 5), (2 vs. 5), and (1 vs. 5), respectively. The one-tailed *P*-values of all comparisons are listed at the bottom of Figure 9. From the figure, it is observed that every iteration improves the game equilibrium payoff with different extents, while the equilibrium result of the last iteration (5) is significantly larger than those of all the previous iterations.

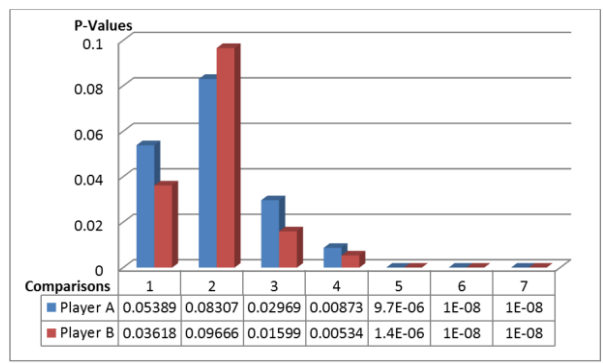

**Fig. 9** *P*-values for comparisons of solution payoffs between iterations

Lastly, experiment results on game stability issues are provided in Table 9. To ensure the steady state, each player was deciding its strategies repeatedly for an extremely large amount of times (e.g. 2000 times/steps in our study). From Table 9, a decreasing trend of instable area is observed through Iterations 1 to 5 (from 26.17% to 14.84%). Considering the stability set from Iterations 1 to 5, the asymptotic stability area increases from 2.34% to 69.53%, and the marginal stability area decreases from 71.48% to 15.63% (Iteration 4 is an abnormal case and needs further investigation). A larger stable area brings a greater portion of points that can eventually converge to the game equilibrium or its acceptable tolerance region. So, the players or game analyst will have an increased confidence to believe that the calculated equilibrium could be achieved.

**Table 9** Comparisons of profile stability under tolerance $\varepsilon = 1500$

| Iteration | Ratio of AS profiles | Ratio of MS profiles | Ratio of instable profiles |
|---|---|---|---|
| 1st | 2.34% | 71.48% | 26.17% |
| 2nd | 14.84% | 57.81% | 27.34% |
| 3rd | 21.09% | 58.20% | 20.70% |
| 4th | 10.55% | 76.56% | 12.89% |
| 5th | 69.53% | 15.63% | 14.84% |

*AS: Asymptotic Stable, MS: Marginal Stable

In addition to the game-theoretic analysis, the proposed simulation framework can be used to help the company managers gain useful insights through comparative analysis. For example, Figure 10 summarizes the simulation state comparisons between equilibriums of Iterations 1 and 5. Note that the horizontal axis in all figures is the simulation run length. It is observed that the warm-up period takes roughly 40~50 days, so the simulation replication length has been set as 100 days (horizontal axis) to reach the system steady state. In addition, the random noises and disturbances are intentionally created to test how both players perform. As observed in Figure 10, although the averages of all the outputs are almost identical, the simulation steady state of game equilibrium in Iteration 5 (right figures in Figure 10) in general is more stable and involves less variations than the one in Iteration 1 (left figures in Figure 10) given the same amount of noises and disturbances. The weak dominance in Iteration 1 is more sensitive and may change between the two players over time depending on the disturbances. The changing trend tends to last long, and the changing amount tends to accumulate high before company takes appropriate actions to compensate. Under the strict equilibrium, the dominance is shared by the two players and is not quite sensitive to the disturbances.

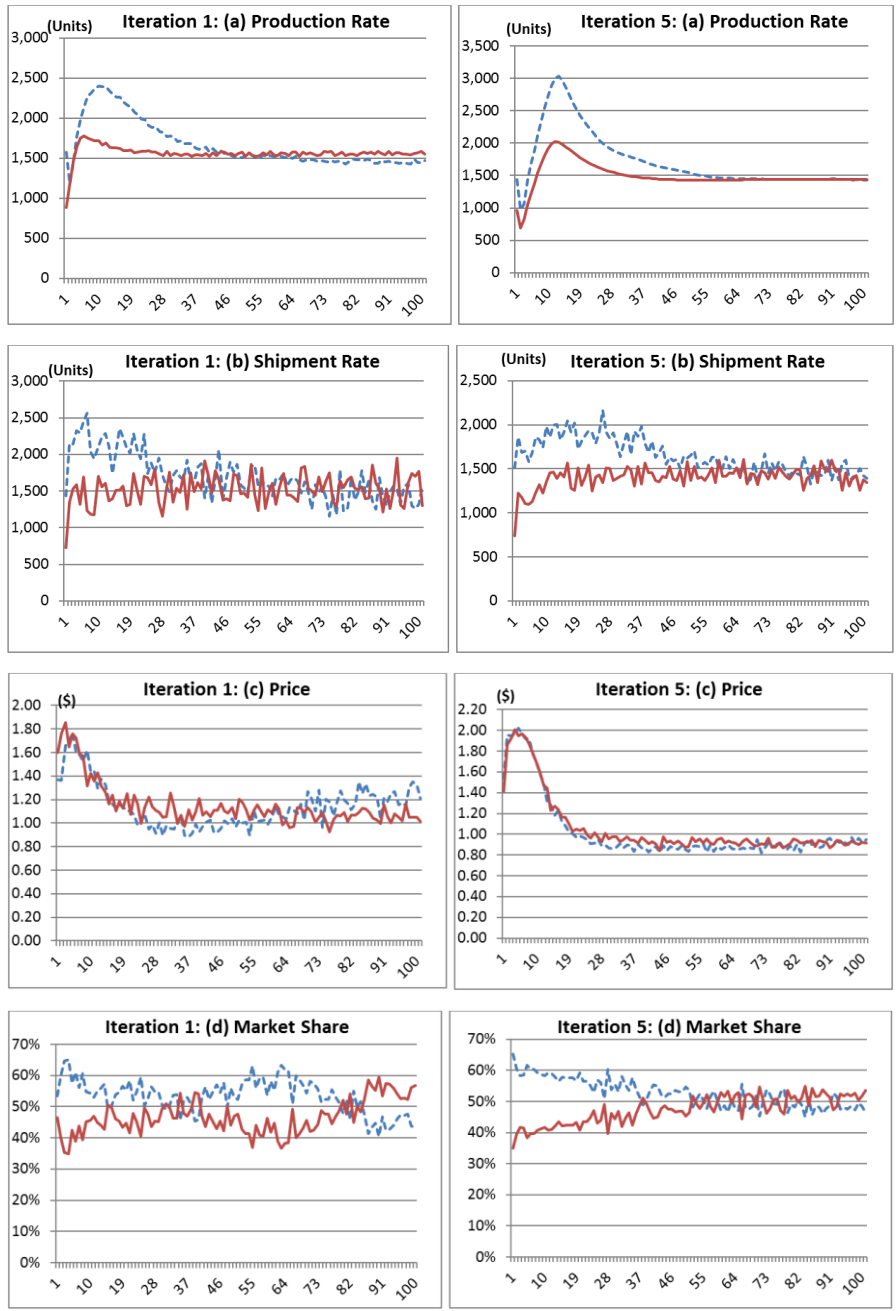

**Fig. 10** Simulation steady state comparisons of game equilibrium on player A (dotted) and player B (straight) for different aspects between Iterations 1 (left) and 5 (right)

### 5.3 Experiment Summary

In summary, given the duopoly case study scenario discussed in Sect. 5.1, as the iteration proceeds in our experiments, the following experimental results have been found:

1. The estimated solution payoff can reach the true solution payoff closer, which enhances the accuracy of equilibrium results.
2. The game solution has moved from a weak to a strict equilibrium, which improves the quality of game equilibrium.
3. The estimated payoffs for both players increase, which provides a better win-win situation to the game.
4. The asymptotic and marginal stable profiles with respect to the game equilibrium are found to increase, which enhances the game stability.

## 6. Conclusions and Future Directions

In this chapter, we proposed a novel hybrid simulation model which integrates agent-based simulation for consumer market activities and system dynamics model for duopoly companies' supply chain operations. Based on the proposed model, we developed a novel GSA procedure, which involve various components such as strategy refinement, data sampling, game solving, and performance evaluation to resolve the simulation-based empirical game. Then, experiments are conducted, where soft drink duopoly scenarios are considered involving different decision variables and experimental iterations. Experiment results have successfully demonstrated

- effectiveness of proposed simulation framework in terms of integrating supply chain operations, marketing activities, and estimating the player strategic movement
- effectiveness of proposed GSA procedure in terms of achieving reduced estimated errors, improvement, robustness, and stability for game equilibriums

Future researches will focus on the following aspects. A variety of simulation scenarios are in the list to further test the scalability issues of the proposed simulation testbed with the GSA procedure. A mathematical proof for the effectiveness and convergence of the proposed GSA procedure will enhance the practicability issue and help to adapt the approach depending on distinct conditions.

## References

Asano M, Iryo T, Kuwahara M (2010) Microscopic pedestrian simulation model combined with a tactical model for route choice behavior. Transp Res C 18: 842-855

Adjali I, Dias B, Hurling R (2005) Agent based modeling of consumer behavior. In: Proceedings of the 2005 North American Association for Computational Social and Organizational Science Annual Conference, University of Notre Dame, Notre Dame, IN, USA

Albert R, Barabasi A (2002) Statistical mechanics of complex networks. Rev of Mod Phys 74: 47-97

Bertrand J (1883) Book review of theorie mathematique de la richesse sociale and of recherches sur les principles mathematiques de la theorie des richesses. J Savants 67: 499–508.

Collins J, Arunachalam R, Sadeh N, Eriksson J, Finne N, Janson S (2004) The supply chain management game for the 2005 trading agent competition. Carnegie Mellon University, Pittsburgh, PA, USA

Cournot competition http://en.wikipedia.org/wiki/Cournot_competition. Accessed 23 Jun 2012

CMO Council CMO Council State of Marketing, Intentions and Investments for 2010. http://www.deloitte.com/assets/Dcom-UnitedStates/Local%20Assets/Documents/us_consulting_CMOCouncil_050510.pdf Accessed 25 November 2012

Esmaeili M, Aryanezhad M, Zeephongsekul P (2009) A game theory approach in seller-buyer supply chain. Eur J Oper Res 195: 442-448

Hong I, Hsu H, Wu Y, Yeh C (2008) Pricing decision and lead time setting in a duopoly semiconductor industry. In: Proceedings of the Winter Simulation Conference (WSC), 2229-2236

Jager W, Janssen M A, Vlek C A J (1999) Consumats in a commons dilemma: testing the behavioral rules of simulated consumers. COV Report No. 99-01, Centre for Environment and Traffic Psychology, University of Groningen

Mockus J (2010) On simulation of optimal strategies and Nash Equilibrium in the financial market context. J Glob Optim 48: 129-143

Jordan P R, Vorobeychik Y, Wellman M P (2008) Searching for Approximate Equilibria in empirical games. In: Proceedings of 7th International Conference on Autonomous Agents and Multiagent Systems (AAMAS), 1063-1070

Jordan P R, Schvartzman L J, Wellman M P (2010) Strategy exploration in empirical games. In: Proceedings of 9th International Conference on Autonomous Agents and Multiagent Systems (AAMAS), 1131-1138

Kiekintveld C, Wellman M P, Singh S. (2006) Empirical game-theoretic analysis of chaturanga. In: AAMAS-06 Workshop on Game-Theoretic and Decision-Theoretic Agents

Kotler P, Keller K L (2007) A framework for marketing management. 3rd edn. Pearson Prentice Hall, Upper Saddler River

Min H, Zhou G (2002) Supply chain modeling: past, present and future. Comput Ind Eng 43: 231-249

Morris B (1987) Coke vs. Pepsi: cola war marches on. Wall Str J 33

Naik P A, Raman K, Winer R S (2005) Planning marketing-mix strategies in the presence of interaction effects. Mark Sci 24: 25–34

North M J, Macal C M, St Aubin J, Thimmapuram P, Bragen M, Hahn J, Karr J, Brigham N, Lacy M E, Hampton D (2010) Multiscale agent-based consumer market modeling. Complex 15: 37-47

Parlar M Wang Q (1994) Discounting decisions in a supplier-buyer relationship with a linear buyer's demand. IIE Trans 26: 34-41

Poropudas J, Virtanen K (2010) Game-theoretic validation and analysis of air combat simulation models. IEEE Trans on Syst Man Cybern A 40: 1057-1070

Reeves D M (2005) Generating trading agent strategies: analytic and empirical methods for infinite and large games. Ph.D. Dissertation, the University of Michigan

Song F, Jing Z (2010) The analysis and system simulation of price competition in a duopoly market. In: Proceedings of Management and Service Science (MASS), 1-4

Sterman J D, (2000) Business dynamics, systems thinking and modeling for a complex world. McGraw-Hill, New York

Sureka A, Wurman, P R (2005) Using Tabu best-response search to find pure strategy Nash Equilibria in normal form games. In: 4th International Joint Conference on Autonomous Agents and Multiagent Systems, 1023–1029

Szidarovszky F, Bahill A T (1998) Linear Systems Theory. 2nd Edn, CRC Press, Boca Raton.

Unsal H H, Taylor J E, (2011) Modeling inter-firm dependency: game theoretic simulation to examine the holdup problem in project networks. J Constrn Eng Manag 137: 284–293

Venkateswaran J, Son Y J (2007) Effect of information update frequency on the stability of production-inventory control systems. Int J Prod Econ 106: 171-190

Vorobeychik Y, Wellman M P, Singh S (2007) Learning payoff functions in infinite games. Mach Learn 67: 145–168

Vorobeychik Y (2008) Mechanism design and analysis using simulation-based game models. Ph.D. Dissertation, the University of Michigan

Vorobeychik Y (2010) Probabilistic analysis of simulation-based games. ACM Trans Model Comput Simul Vol 20 No 3 Article 16

Walsh W E, Parkes D C, Das R (2003) Choosing samples to compute heuristic-strategy Nash Equilibrium. In: AAMAS Workshop on Agent Mediated Electronic Commerce V (AMECV) Melboune, Australia

Wang Q, Wu Z (2000) Improving a supplier's quantity discount gain from many different buyers. IIE Trans 32: 1071-1079

Wellman M P (2006) Methods for empirical game-theoretic analysis. In: AAAI'06 Proceedings of the 21st National Conference on Artificial Intelligence 2: 1552-1555

Yoshida T, Hasegawa M, Gotoh T, Iguchi H, Sugioka K, Ikeda K (2007) Consumer behavior modeling based on social psychology and complex networks. In: E-Commerce Technology and the 4th IEEE International Conference on Enterprise Computing, E-Commerce and E-Services, 493-494

Yu Y, Liang L, Huang G (2006) Leader-follower game in vendor-managed inventory system with limited production capacity considering wholesale and retail prices. Int J Logist Res Appl 9: 335-350

Zhang T, Zhang D (2007) Agent-based simulation of consumer purchase decision-making and the decoy effect. J Bus Res 60: 912-922

Zhang X, Huang G (2010) Game-theoretic approach to simultaneous configuration of platform products and supply chains with one manufacturing firm and multiple cooperative suppliers. Int J Prod Econ 124: 121-136